\documentclass[12pt]{iopart}
\usepackage[usenames,dvipsnames]{xcolor}
\usepackage{graphicx}  
\usepackage{epstopdf}  
\usepackage{overpic}   
\usepackage{dcolumn}   
\usepackage{bm}        %
\usepackage{amssymb}   %
\usepackage{textcomp}
\usepackage{color}   
\begin{document}

\title{Two-Plasmon-Decay Instability Stimulated by a Normal- and Large-Angle-Incidence Laser Pair}

\author{C.-W. Lian $^{1}$, Y. Ji$^1$, R. Yan$^{1,3*}$, J. Li$^{2\#}$, S.-H. Cao$^{4}$, C. Ren$^{4}$, L.-F. Wang$^5$, Y.-K. Ding$^5$, and J. Zheng$^{2,3}$}

\address{$^1$Department of Modern Mechanics, University of Science and Technology of China, Hefei, Anhui 230026, China \\
$^2$Department of Plasma Physics and Fusion Engineering, University of Science and Technology of China, Hefei 230026, China \\
$^3$Collaborative Innovation Center of IFSA, Shanghai Jiao Tong University, Shanghai 200240, China \\
$^4$Department of Mechanical Engineering, University of Rochester, Rochester, New York 14627, USA\\
$^5$Institute of Applied Physics and Computational Mathematics, Beijing 100088, China}
\ead{* ruiyan@ustc.edu.cn, \quad \# junlisu@ustc.edu.cn}
\vspace{10pt}
\date{\today}
\begin{abstract}
	The two-plasmon-decay instability (TPD) is a critical target preheating risk in direct-drive inertial confinement fusion. In this paper, TPD collectively driven by a normal-incidence laser beam (Beam-N) and a large-angle-incidence laser beam (Beam-L) is investigated via particle-in-cell simulations. Significant TPD growth is found able to develop in this regime at previously unexpected low laser intensities if the intensity of Beam-L exceeds the large-angle-incidence threshold. Both beams contribute to the growth of TPD in a ``seed-amplification" manner where the absolute instability driven by Beam-L provides the seeds that get convectively amplified by Beam-N, making TPD energetically important and causing significant pump depletion and hot electron generation.
\end{abstract}

%
%
%
%
%

\section{Introduction}
As the milestone on inertial confinement fusion (ICF) ignition has been achieved\cite{Zylstra2022,Abu2022,Betti2023} in the National Ignition Facility (NIF), higher energy gain is being pursued by the ICF community in the next phase. The direct-drive schemes, in which the high-intensity lasers directly deposit energy to an ICF target, generally have high laser-to-target energy coupling efficiency and are considered as candidates for high-gain fusion approaches.\cite{Craxton2015} Direct laser interaction with target is not only a key process in the conventional central-hot-spot direct-drive scheme, but is also utilized in one or multiple phases of novel ignition schemes such as the shock ignition\cite{Betti2007}, the hybrid ignition\cite{He2016}, and the double-cone ignition\cite{Zhang2020}. 

One of the critical challenges in the direct-drive ICF is the harmful impacts due to the laser plasma instabilities (LPIs), including stimulated Raman Scattering (SRS), stimulated Brillouin Scattering (SBS), cross beam energy transfer (CBET), and two-plasmon decay (TPD) instability, occurring in the corona plasmas ablated off from the target shell. As the laser propagates through a direct-drive corona plasma, it experiences a broad range of electron density from vacuum up to the critical density ($n_{cr}$) beyond which a laser is not able to propagate. In this scenario, the quarter-critical density ($n_{cr}/4$) areas, where TPD and the absolute SRS can be parametrically stimulated, are usually located in a less collisional plasmas consisting of low-Z ablator materials and therefore are of critical concern. The LPI processes near $n_{cr}/4$ are of less concern in an indirect-drive hohlraum since the $n_{cr}/4$  areas are usually located in a high-Z gold plasma where the heavy collisional damping inhibits TPD or SRS growths there. LPIs can not only cause direct energy loss through the processes involving scattered lights, i.e., SRS, SBS, and CBET, but also cause fuel preheating threats through the energetic (hot) electrons produced via TPD and SRS which involve electron plasma waves(EPWs) that are able to accelerate electrons.

Through TPD a laser parametrically decays into a pair of the so-called daughter EPWs. TPD is considered as a critical LPI impacting the implosion performance in direct drive due to its low threshold\cite{Liu1976,Simon1983,Afeyan1997} and hot electron generation\cite{Smalyuk2008HE,Yan2012}. TPD has been identified as the primary source of hot electrons which can preheat the fuel\cite{Craxton2015,Smalyuk2008HE} and compromise the implosion performance on the OMEGA direct-drive facility.

While the modern ignition facilities generally use multiple overlapping laser beams, much of the knowledge on TPD was firstly established in simplified single-beam regimes.\cite{Liu1976, Simon1983, Yan2012, Langdon1979, Baldis1981, DuBois1995,Russell2001,Yan2009,Yan2010}
The threshold for TPD growth in an inhomogeneous plasma with a linear density profile in a normal-incidence regime [i.e., the laser propagation direction is along the electron density ($n_e$) gradient $\nabla n_e$] was obtained\cite{Simon1983} as 
\begin{equation}
	\label{eq:simon-eta}
	\eta \equiv L_{\mu}\lambda_\mu I_{14}/(82T_{keV})>1,
\end{equation}
where $L_\mu$ is the electron density  scale length $L_n$ at $n_{cr}/4$ in the unit of microns, $\lambda_\mu$ is the laser wavelength in vacuum ($\lambda_0$) in microns, $I_{14}$ is the laser intensity ($I_{0}$) in the unit of $10^{14}\ \mathrm{W/cm}^2$, and $T_{keV}$ is the electron temperature ($T_e$) in $\ \mathrm{keV}$. In references \cite{Yan2009,Yan2010} two different types of modes, absolute and convective, were identified via fluid simulations based on their distinguished growth behaviors in a linear system. Absolute modes grow exponentially with time while convective modes have limited factors of amplification on the initial seeds. However, the convective modes were found energetically important in the simulations\cite{Yan2009} due to the nonlinearity involving ion dynamics\cite{Yan2012, DuBois1995,Russell2001,Yan2009}therefore both absolute and convective modes saturate forming quasi-static broad EPW spectra up to the Landau cutoff in the $k$ space. Broad spectra of the $3 \omega_0/2$ emission associated with the TPD EPWs have been consistently observed in OMEGA experiments\cite{Seka2009}, where $\omega_0$ is the laser frequency.

It has been well recognized that TPD can be collectively driven by multiple overlapping lasers. The total energy in hot electrons was shown experimentally to scale with the overlapped intensity on OMEGA.\cite{Stoeckl2003} The threshold condition of Eq.(\ref{eq:simon-eta}) which has been largely verified by the fluid simulations\cite{Yan2010} also works empirically well predicting the onset of TPD in OMEGA experiments if using the overlapped intensity to calculate $\eta$.\cite{Seka2009} These works evidenced that TPD is highly unlikely to be independently driven by each individual laser beam. Deliberate experiments were then carried out to investigate the growths of the shared daughter plasma waves (often referred to as the common waves) contributed by multiple laser beams.\cite{Michel2012, Michel2013} It was evidenced that laser beams can drive the common waves in the region of wave numbers bisecting the beams and the convective gain is proportional to the overlapped laser intensity.\cite{Michel2012} Moreover,  the hot electron fraction was found to increase exponentially with the overlapped laser intensity in the experiments under various beam configurations. \cite{Michel2013} Three dimensional (3D) simulations further showed that the threshold for the absolute TPD modes driven by multiple beams is significantly lower than the expected single-beam value. \cite{Zhang2014}

An important factor that can contribute to the low threshold found in reference \cite{Zhang2014} comes from the oblique incidence angles. The single-beam threshold intensity of TPD under moderate incidence angles was found to be lowered by a factor of $\sim \cos^2 \phi$.\cite{Wen2016}, where $\phi$ is the angle between the local laser wave vector $\mathbf{k}_{loc}$ and $\nabla n_e$ near $n_{cr}/4$. 
Large incidence angles, however, lower the absolute TPD threshold even more drastically. \cite{Lian2022,Zhou2023} To clarify, the incidence angle ($\theta$) is referred to as the angle between $\mathbf{k}_{0}$ and $\nabla n_e$ in this paper, where $\mathbf{k}_{0}$ is the laser wave vector in vacuum with the magnitude $k_0 \equiv 2\pi/\lambda_0 = \omega_0/c$ and $c$ is the speed of light in vacuum. The regime where $\theta$ is close to $60^\circ$ or even larger is considered as the large-angle-incidence regime.
Based on theoretical modeling and fluid simulation using \textit{LTS} \cite{Yan2010}, Lian et al. \cite{Lian2022} proposed a threshold  for the absolute TPD growth in the large-angle-incidence regime with $\theta = 60^\circ$:
\begin{equation}
\xi \equiv L_{\mu}^{4/3} \lambda _{\mu}^{2/3} I_{14}/(3T_{keV})>1,
\label{eq:xi}
\end{equation}
which was also found consistent with the PIC simulations\cite{Lian2022}. This large-angle-incidence TPD threshold was shown to be two orders of magnitude lower than that of the normal incidence with ICF-relevant parameters.\cite{Lian2022}
Moreover, the simulations on TPD induced by multiple beams all with  $\theta\approx 60^\circ$ also exhibited low threshold and common-wave processes which largely depend on the lasers' polarization.\cite{Zhou2023}

The very low TPD threshold in the large-angle-incidence regime brings attention to TPD in a sophisticated laser-plasma system where virtually all incidence angles are present and a portion of plasma can be simultaneously irradiated by both small- and large-angle-incidence beams. Whether the large-angle-incidence TPD which can be driven at low laser intensities can cause destabilizing effects in the whole system is an open question. 

In this paper, we study TPD in a simpler but typical ``N-L'' two-beam system composed of a Normal-incidence beam (Beam-N) and a Large-angle-incidence beam (Beam-L) via numerical simulations. The schematic of an N-L system in direct drive is illustrated in figure \ref{fig:scheme}(a). Although each beam in direct drive is generally pointed to the target center, the off-axis part of the laser with weaker intensity than the peak value can glance at the edge part of the corona plasma surrounding the target. Therefore the zone inside the box in figure \ref{fig:scheme}(a) can be recognized to be irradiated by a higher intensity Beam-N and a lower-intensity Beam-L. N-L systems typically exist in many direct-drive schemes such as the conventional central-hot-spot ignition scheme \cite{Craxton2015}, the shock ignition\cite{Betti2007,Perkins2009}, and the double-cone ignition\cite{Zhang2020}. The rest of this paper is organized as follows: In Sec. II, we describe the simulation set-up. The development of TPD in an N-L configuration is investigated in Sec. III and the generation of hot electrons is discussed in Sec. IV. A summary is presented in Sec. V.

\section{Simulation configurations}
\begin{figure}[!th]
	\begin{center}
		\includegraphics[width=3.375in]{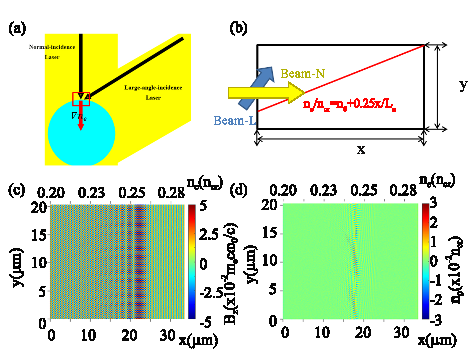}
		\caption{(color online).(a) Schematic of the N-L configuration in direct drive with the red box indicating the simulation domain; (b) schematic of the simulation setup; (c) $B_z$ of Case I at $t = 1.0 \ \mathrm{ps}$; (d) the density perturbation of electrons ($n_p$) of Case I at $t = 2.0 \ \mathrm{ps}$.  In (d), a Gaussian filter $ G(x, y) = e^{-(x^2 + y^2)/2\sigma^2}/2\pi\sigma^2 $ with $\sigma = 2$  is applied to reduce the background noises.}
		\label{fig:scheme}
	\end{center}
\end{figure}

A series of Particle-in-Cell(PIC) simulations are performed using a full PIC code $OSIRIS$ \cite{Fonseca2002} to investigate the LPI processes in the boxed zone in figure \ref{fig:scheme}(a). The schematic of a typical simulation box is illustrated in figure \ref{fig:scheme}(b). The parameters are chosen relevant to OMEGA and SG direct-drive experiments. The plasma has a linear electron density profile as $n_e(x) = [0.20 + 0.25x/L_n] n_{cr} $ and a uniform electron temperature $T_e = 3  \ \mathrm{keV}$, where the electron density scale length is set as $L_n = 100  \ \mathrm{\mu m}$ in all of the simulations cases in this paper.  The ions are composed of fully ionized CH (plastic) with the number ratio $1:1$ and the temperature $T_i = 1.5 \ \mathrm{keV}$ unless otherwise noted. The electron-ion collisions are included.
Two laser beams, ie., Beam-N and Beam-L, are launched from the left side with a rising time $0.2 \ \mathrm{ps}$ and at the incident angles $\theta = 0^\circ$ and $58.7^\circ$, respectively. Different combinations of the laser intensities of Beam-N ($I_N$) and Beam-L ($I_L$) are explored and the detailed parameters are listed in TABLE \ref{tab:param}. Since Beam-L is abstracted from the off-axis part of an actual laser beam [see figure \ref{fig:scheme}(a)], it is physical to set $I_L$ much weaker than $I_N$. Both beams are linearly polarized in the $x-y$ plane. 

The simulation box size is $34 \ \mathrm{\mu m}\times 20 \ \mathrm{\mu m}$ with a grid of $3000 \times 1820$, yielding the grid size $\Delta x = \Delta y = 0.2 c/\omega_0$.
For each species (electrons and ions), 100 particles per cell are used. The simulation domain is periodic in $y$. 
Open boundary conditions are applied to the electromagnetic fields in the $x$ direction.  The ``thermal-bath" boundary conditions, which reflect the particles reaching the boundary and assign them with new random velocities following the initial Maxwellian distribution, are applied to the particles in the $x$ direction.

\section{Growth of TPD in the N-L system}
\begin{figure}[th]
\begin{center}
\includegraphics[width=3.375in]{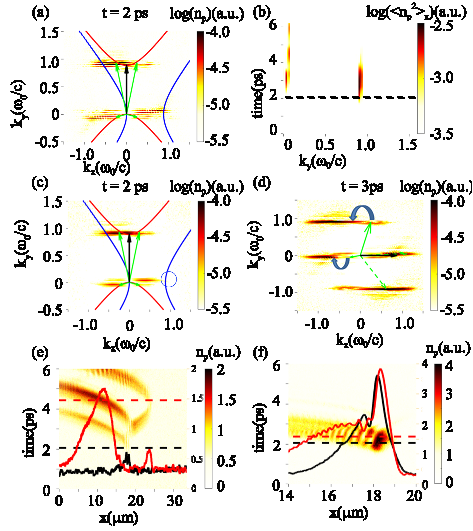}
\caption{(color online). (a) $n_p$ of Case I at $t = 2.0 \ \mathrm{ps}$ in the $k_x-k_y$ space, with the wave vectors of the EPWs (Beam-L near $n_e = 0.25 n_{cr}$) marked by the green (black) arrows. The red curves represent the maximum growth curves of TPD in the homogeneous plasma of Beam-L, while the blue curves depict the maximum growth curves of Beam-N. (b) Time evolution of $\langle n_p^2\rangle_x$ in the $k_y$ space in Case VI. Here, the black dashed line at $t = 2.0 \ \mathrm{ps}$ indicates the switch between Beam-L and Beam-N. (c) $n_p$ of Case VI at $t = 2.0 \ \mathrm{ps}$ in the $k_x-k_y$ space with the distribution difference compared to Case I mark by the blue circle. (d) $n_p$ of Case VI at $t = 3.0 \ \mathrm{ps}$  in the $k_x-k_y$ space, where the green (black) arrows are the wave vectors of the EPWs (Beam-N near $n_e =0.25 n_{cr}$). Here, the green solid (dashed) arrows represent the wave vectors of EPWs excited before(after) Beam-N incidence and the blue arrows represent the convective amplification process of the EPWs. (e) Time and space evolution of $n_p$ with $k_y = 0.07 \omega_0/c$ and the $n_{p}$ along the $x$ direction at $t = 2 \ \mathrm{ps}$ (black line) and $t = 4.5 \ \mathrm{ps}$ (red line) in Case VI. (f) Time and space evolution of $n_p$ with $k_y = 0.9 \omega_0/c$ and $n_p$ along the $x$ direction at $t = 2 \ \mathrm{ps}$ (black line) and $t = 2.2 \ \mathrm{ps}$ (red line) in Case VI.}
\label{fig:sep}
\end{center}
\end{figure}

The simulation parameters are chosen such that $I_N$ is well below the threshold namely $\eta \le 0.71$ such that the normal-incidence Beam-N is not intense enough to excite TPD by itself. Beam-L is close to glancing incidence near the turning point approximately located at $n_e = 0.27 n_{cr}$. The laser fields in Case I are shown in figure \ref{fig:scheme}(c). In this case $\xi = 13 \gg 1$, indicating that Beam-L is well above the TPD threshold and able to drive absolute growth in the large-incidence-angle regime if shined alone. It should be noted that the ``overlapping $\eta$'' ($\eta_{all}$) calculated by substituting $I_{all} \equiv I_N + I_L$ into Eq.(\ref{eq:simon-eta}) are also below unity in all of the cases, as listed in TABLE \ref{tab:param}. 
A snapshot of the electron density fluctuations ($n_p$) associated with EPWs in Case I is illustrated in figure \ref{fig:scheme}(d), and the time and space evolution of $\langle  n_p^2 \rangle_y (x,t)$ is illustrated in figure \ref{fig:sep}(a). Here the bracket $\langle ... \rangle_y$ represents averaging over $y$.  It is shown that EPWs with notable amplitude are driven near 0.25$n_{cr}$ and last throughout the simulation. Those EPWs are attributed to TPD instead of SRS, since virtually no SRS scattered light appears in the spectrum of the $z$ component of magnetic fields ($B_z$) which is associated with the lights. The presence of the large-incidence-angle Beam-L is identified as the key factor enabling TPD growth in Case I, as no TPD is observed if we redistribute all the intensity from Beam-L to Beam-N while maintaining the same $I_{all}$ (see Case III of TABLE \ref{tab:param}).

\begin{figure}[th]
	\begin{center}
		\includegraphics[width=3.375in]{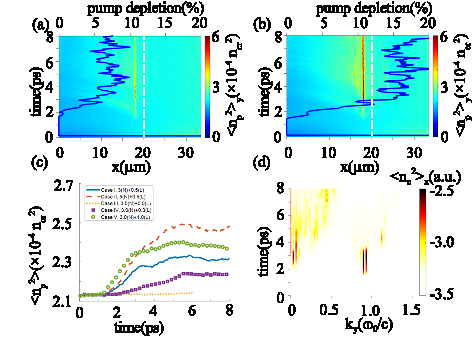}
		\caption{(color online).(a) Time and space evolution of $\langle n_p^2\rangle_y$ in Case I with the pump depletion of Beam-N marked by the blue solid line and $0.25 n_{cr}$ marked by the white dashed line. Here, the bracket $\langle ... \rangle_y$ represents averaging over $y$. (b) Time and space evolution of $\langle n_p^2\rangle_y$ in Case II; (c) Time evolution of $\langle n_p^2\rangle$ in Cases I, II, III, IV and V, where $\langle n_p^2\rangle$ is the averaged value of $n_p^2$ over the whole simulation domain; (d) time evolution of $\langle n_p^2\rangle_x$ in the $k_y$ space in Case I. Here, $\langle n_p^2 \rangle_x$ represents $n_p^2$ averaging over $x$ after applying the Fourier transform on $n_p$ along the $y$ direction.} 
		\label{fig:simult}
	\end{center}
\end{figure}

Once TPD is able to grow, it can be energetically important and cause significant pump depletion in the N-L system. Figure \ref{fig:simult}(a) demonstrates the pump-depletion fraction of Beam-N in Case I. Roughly 7\% of the Beam-N energy is depleted once TPD is stimulated, which evidences that Beam-N also participates in the TPD process. Changing the intensity of either Beam-L or Beam-N is found to change the pump depletion fraction and $\langle n_p^2 \rangle$, which can be used to assess the level of TPD. Here $\langle n_p^2 \rangle$ is the averaged value of $n_p^2$ over the whole simulation domain.
In Case II the intensity of Beam-N is increased to $5 \times 10^{14}\ \mathrm{W/cm}^2$ corresponding to $\eta = 0.71$ which is still lower than the normal threshold, while the intensity of Beam-L is kept the same as in Case I. The pump-depletion of Beam-N rises to roughly 15\%. It should be pointed out that the energy Beam-N loses can be even higher than the input energy of Beam-L.
In Cases III, IV, and V different $I_L$ are used while keeping $I_N$ the same as in Case I to further demonstrate the sensitive dependence of TPD on the fairly low intensity Beam-L. Figure \ref{fig:simult}(c) compares $\langle n_p^2 \rangle$ with different intensities of Beam-N and Beam-L. It is shown that  $\langle n_p^2 \rangle$ saturates to quasi-steady values after a few ps growing. It is also shown that raising either $I_N$ or $I_L$ leads to higher TPD levels and the absence of Beam-L (Case III) completely turns TPD off. 
Both the influence from the low-intensity Beam-L and the significant pump depletion of the high-intensity Beam-N are evidencing that TPD is collectively driven by both beams.

The electron density perturbation in Case I is displayed in the $k_y - t$ space to illustrate the evolution of EPWs from the linear growing stage up to the highly nonlinear stage [see figure \ref{fig:simult}(d)]. The narrow-band EPW spectra concentrated near $k_y \approx 0.9 \omega_0/c$ and $0.07 \omega_0/c$ in the early linear stage become much broader in $k_y$ in the highly nonlinear stage. The modes initially growing in the linear stage are attributed to Beam-L. The dispersion relation for single-beam TPD growing in homogeneous plasmas predicts the well-known fastest-growing hyperbola \cite{Kruer2003} shown by the solid curves in figure \ref{fig:sep}(a), red for Beam-L and blue for Beam-N, respectively. The dominant modes laying on top of the red hyperbola also satisfy the wave vectors' matching condition illustrated by the arrows:
\begin{equation}
	\label{eq:matching-condition}
	\mathbf{k}_{L} = \mathbf{k}_1 + \mathbf{k}_2,
\end{equation}
where $\mathbf{k}_1$ and $\mathbf{k}_2$ (the green arrows) are the wave vectors of the paired daughter EPWs, $\mathbf{k}_{L}$ is the local wave vector of Beam-L (the black arrow), which is propagating along the $y$ direction near its reflecting point. 

The individual roles of the two beams in developing TPD in the N-L system are further demonstrated by separating their time windows in Case VI. In Case VI Beam-L is set to be incident from $0$ to 2 ps while the onset of Beam-N is set on $t = 2$ ps such that they barely overlap in time. All the other simulation parameters are kept essentially the same as Case I, except that the ions are fixed in Case VI to inhibit nonlinear effects brought in by the ion dynamics. The time evolution of the spectrum of $n_p$ is illustrated in figure \ref{fig:sep}(b). It is shown that similar to Case I in linear stage [see figure \ref{fig:simult}(d)] the dominant modes in Case VI are located at $k_y \sim 0.07 \omega_0/c$ and $k_y \sim 0.9 \omega_0/c$, respectively. We then call the modes with $k_y \sim 0.07 \omega_0/c$ the small-$k_y$ modes, the modes with $k_y \sim 0.9 \omega_0/c$ the large-$k_y$ modes. The large-$k_y$ modes are able to grow to a substantial level before the onset of Beam-N, evidencing that the large-$k_y$ modes can be stimulated by the Beam-L alone. During the incidence of Beam-N ($>$2 $\mathrm{ps}$), the amplitudes of the large-$k_y$ modes get further amplified and then decay after 4 ps. The amplification after 2 ps shows that Beam-N transports energy to the large-$k_y$ modes while the decay after 4 ps shows that Beam-N by itself cannot maintain the level, which is a feature of a convective amplification.

The evolution of the large-$k_y$ modes in Case VI can be divided into two phases: excitation by Beam-L and amplification by Beam-N.
The spectra of the EPWs in Case VI in the $k_y-k_x$ space are depicted in figure \ref{fig:sep}(c) at $t = 2$ ps and figure \ref{fig:sep}(d) at $t = 3$ ps to illustrate the TPD matching conditions and the dominant modes' shifting at different phases. 
In the excitation phase [see figure \ref{fig:sep}(c)], the spectrum of Case VI exhibiting both the large-$k_y$ and small-$k_y$ modes are similar to that of Case I [figure \ref{fig:sep}(a)]. It demonstrates that Beam-L itself can stimulate the large-$k_y$ and small-$k_y$ modes simultaneously, which satisfy the wave vectors' matching condition of equation (\ref{eq:matching-condition}) sketched with the solid arrows. Compared to figure \ref{fig:sep}(a), the spectrum in figure \ref{fig:sep}(c) misses a portion near the blue hyperbola due to the lack of Beam-N, as marked by the blue circle. The missing portion can be seen in figure \ref{fig:sep}(d) which shows the dominant modes in the amplification phase with  Beam-N only.  The dominant $k_x$ of both large-$k_y$ and small-$k_y$ modes becomes larger. The dominant modes' shifting can be attributed to two processes: Firstly, by satisfying its own matching conditions, Beam-N can drive and amplify the paired EPWs (marked by the green dashed arrows) of the existing EPWs (marked by the green solid arrows) excited by Beam-L. Secondly, $|k_x|$ of the existing EPWs increase as they propagate towards the lower density, denoted by the blue flow arrows. The propagation of the EPWs and the convective amplification on the path for the large-$k_y$ and small-$k_y$ modes are further illustrated in figure \ref{fig:sep}(e) and (f). 

A set of typical small-$k_y$ modes with $k_y = 0.07 \omega_0/c$ are localized near $x  = 18 \mu$m right below 0.25$n_{cr}$ at $t=2 $ps after the absolute growth in the excitation phase driven by Beam-L only, as shown in figure \ref{fig:sep}(e). The amplitude of $n_p$ along the black dashed line is plotted in the back solid line to show the location of the dominant modes at $t=2 $ps. Then, instead of continuing to grow locally, the small-$k_y$ modes are found to propagate to different locations and get amplified at certain densities. Three distinguishable propagation paths in figure \ref{fig:sep}(e) suggest that at $t=2 $ps the small-$k_y$ modes include a range of $k_x$, leading to different group velocities that sketch different propagation paths on the $t-x$ space. As the small-$k_y$ modes eventually move to lower density due to the electron density gradient, $|k_x|$ increases since an EPW must satisfy the dispersion relation in plasma while $k_y$ keeps constant as the density is uniform along the $y$ direction. The up shift of $|k_x|$ can be seen in figure \ref{fig:sep}(d), as marked by the blue flow arrow on the small-$k_y$ modes. The profile of the small-$k_y$ modes at $t=4.5 $ps along the red dashed line is plotted with the red solid line in figure \ref{fig:sep}(e), demonstrating the convective amplification of the small-$k_y$ modes by Beam-N.

The convective amplification of a typical large-$k_y$ mode with $k_y = 0.9 \omega_0/c$ is shown in figure \ref{fig:sep}(f). The mode is convectively amplified a bit from $t = 2$ ps (black solid line) to $t = 2.2$ ps (red solid line) as the wave package moves forward. 
Compared with the small-$k_y$ modes, the amplification of this mode is much weaker. The difference in the amplification factor between the large-$k_y$ and small-$k_y$ modes can be estimated with the Rosenbluth-like gain presented in reference \cite{Yan2009}:
\begin{equation}
	\pi \Lambda \approx 2.15(1-0.00881T_{\mathrm{keV}} - 0.0470T_{\mathrm{keV}}\tilde{k}^2_y)\eta,
	\label{Eq:Gain}
\end{equation}
where the Rosenbluth amplification factor of a mode with different $\tilde{k}_y$ is  $f(\tilde{k}_y) = \exp[\pi \Lambda]$, $\tilde{k}_y \equiv {k}_y c/\omega_0$ is a dimensionless $k_y$. Landau damping is absent in equation (\ref{Eq:Gain}). The value $f(0.07) = 2.5$ gives the same order of magnitude of the amplification factor as that in the PIC simulation of Case VI, i.e. $f_{sim}(0.07) \approx 1.8$ reads from figure \ref{fig:sep}(c). According to equation (\ref{Eq:Gain}), the amplification factor with larger $k_y$ is smaller, which is qualitatively consistent with the simulation. However, the gain for the mode with $k_y = 0.9 \omega_0/c$ obtained by equation (\ref{Eq:Gain}) is much larger than the simulation results, likely due to the heavier Landau damping of the EPWs which is not taken into account in equation (\ref{Eq:Gain}). 

Overall, the growth of TPD in the N-L system can be recognized as a ``seed-amplification" process where the absolute instability driven by Beam-L provides the seeds that get convectively amplified by Beam-N. The levels of the seeds are mostly determined by Beam-L, while the amplification factors are determined by Beam-N. The two beams contribute to the TPD growth in different ways, both making TPD energetically important in the N-L system, even after the saturation due to the nonlinear effects including ion motions.

\begin{figure}[th]
\begin{center}
\includegraphics[width=3.375in]{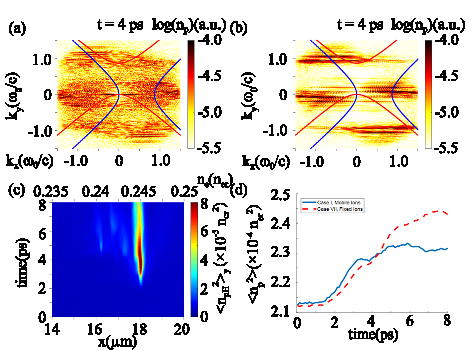}
\caption{(color online). (a) $n_p$ of Case I with mobile ions at $t = 4.0 \ \mathrm{ps}$ in the $k_x-k_y$ space with the maximum growth curves of TPD in homogeneous plasma of Beam-L (Beam-N) marked by the red (blue) curves; (b) $n_p$ of Case VII with fixed ions at $t = 4.0 \ \mathrm{ps}$ in the $k_x-k_y$ space; (c) time evolution of $\langle n_{pH}^2\rangle_y$ in Case I, where $n_{pH}$ is the density perturbation of H ion.}
\label{fig:ion}
\end{center}
\end{figure}

The EPW spectrum in the nonlinear stage at 4 ns is much broader [figure \ref{fig:ion}(a)] than that in the linear stage [figure \ref{fig:sep}(a)] of Case I. Spectrum broadening was also found in the simulations on normal-incidence TPD.\cite{Yan2012, Yan2009}
The ion dynamics are recognized as a key factor in broadening the EPW spectrum. Case VII with immobile ions and virtually the same other configurations as Case I is performed to show the role of the ion motions. Case VII demonstrates a much narrower EPW spectrum than Case I, as shown in figure \ref{fig:ion}. It was well known that the ion density fluctuations are driven by the ponderomotive pressure of the EPWs and are correlated with the level of EPWs.\cite{Yan2009} In Case I where both beams are shining from the very beginning of the simulation, the ion density fluctuations are gradually driven up and mostly localized in a fairly narrow region near 1/4 $n_{cr}$, where the absolute modes are located, shown in figure \ref{fig:ion}(c).

The ion motions also contribute to the TPD saturation and limit the EPWs level. Figure \ref{fig:ion}(d) compares the time evolution of $\langle n^2_{p} \rangle$, which quantifies the EPWs level, of Case VII with fixed ions and of Case I with mobile ions. It is shown that TPD in Case I saturates at a lower level after $t = 3$ ns when the ion density fluctuations are significant (see figure \ref{fig:ion}(c)). It was demonstrated experimentally that the ions can saturate TPD via the Langmuir decay instability \cite{Follett2015}, where a primary TPD EPW decays into an ion acoustic wave and a new EPW. In contrast, TPD in Case IV saturates later and at a higher level, likely due to the convective amplification limits\cite{Yan2010, Rosenbluth1973} and/or other kinetic saturation mechanisms limiting the EPWs amplitudes. It should be pointed out that including ion motions is critical to correct modeling on TPD saturation, which largely determines the EPWs features and the hot electron generation in the nonlinear stage.

\section{Hot electron generation}
The hot electron generation is of most concern since the hot electrons generated by TPD are recognized as a significant target preheating risk. Usually, the hot electrons whose kinetic energy is above 50 keV are considered most threatening and therefore paid most attention in our simulations.
Both the forward and backward moving instantaneous energy fluxes carried by the hot electron above 50 keV are monitored in the simulations at the right and left boundaries of the simulation box as $\alpha_{R}$ and $\alpha_{L}$, respectively. Here $\alpha_{R}$ and $\alpha_{L}$ have been renormalized to the incident laser energy flux including both beams. Then $\alpha_{50}
\equiv \alpha_{L} + \alpha_{R}$ can be recognized as the energy conversion fraction from laser to the hot electrons $>$50 keV.

\begin{figure}[th]
\begin{center}
\includegraphics[width=3.375in]{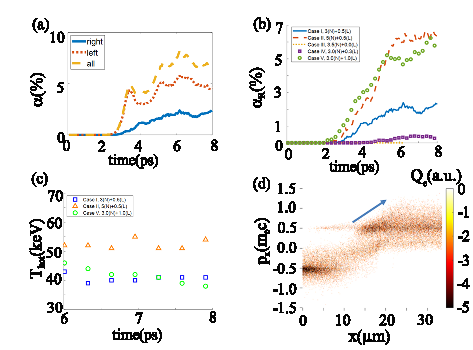}
\caption{(color online). (a) Time evolution of $\alpha$ for the left (red), right (blue) boundaries and both sides (orange) in Case I. Here, $\alpha$ represents the instantaneous energy flux carried by hot electrons ($\geq 50 \ \mathrm{keV}$) monitored in the boundaries and has been normalized to the incident laser energy flux including both beams. (b) $\alpha_{R}$ in Cases I, II, III, IV and V. Here, $\alpha_{R}$ is the $\alpha$ for the right boundary and represents the hot electrons moving to the higher density region. (c) The fitting temperatures of the hot electrons ($T_{hot}$) in Cases I, II and V; (d) the charge density distribution of the hot electrons in the $p_x-x$ phase space at $t = 6 \ \mathrm{ps}$ in Case I. The staged acceleration of electrons is sketched with the blue arrow.}  
\label{fig:hot}
\end{center} 
\end{figure}

From the time evolution of $\alpha$'s in Case I as shown in figure \ref{fig:hot}(a),  substantial laser energy can be transferred to the hot electrons, mostly in the nonlinear stage after $t=3$ ps. $\alpha_{50}$ increases sharply at about $t=3$ ps and reaches a notable level $\sim 7 \%$ later in the nonlinear stage, which is correlated with the result that the pump depletion of Beam-N reaches high levels after about 3 ps as plotted in figure \ref{fig:simult} (a). This correlation evidences that the majority of the hot electron energy comes from Beam-N.  $\alpha_{R}$ is of more concern as it moves forward to the high density side toward the center of an ICF target, causing direct preheating.  $\alpha_{R}$ is found smaller than $\alpha_{L}$ as shown in figure \ref{fig:hot}(a) yet is still alerting as $\alpha_{L} \sim 2 \%$  is much larger than the tolerable $\sim 0.1\%$ hot-electron preheating in direct drive \cite{Craxton2015}.

Increasing the intensity of either Beam-N or Beam-L is found to enhance the hot electron fluxes in the N-L system, as demonstrated by figure \ref{fig:hot} (b). The intensity of Beam-N increases from $I_N = 3\times 10^{14} \mathrm{W/cm^2}$ in Case I to $I_N = 5\times 10^{14} \mathrm{W/cm^2}$ in Case II, causing $\alpha_{R}$ to increase from $2\%$ to $6\%$ in the nonlinear stages after $t=5$ ps.  Increasing the intensity of Beam-L $I_L = 5\times 10^{13} \mathrm{W/cm^2}$ in Case I to $I_L = 1\times 10^{14} \mathrm{W/cm^2}$ in Case V causes $\alpha_{R}$ to rise from $2\%$ to about $5\%$, while lowering $I_L$ down to $I_L = 3\times 10^{13} \mathrm{W/cm^2}$ in Case IV reduces  $\alpha_{R}$ down to $0.2\%$.
These trends are evidencing that enhancing either  ``amplification'' or  ``seed'' in the N-L system would enhance the hot electron generation and vice versa. It is also shown in figure \ref{fig:hot}(b) that the absence of Beam-L in Case III ($I_L =0 $) completely suppresses TPD and produces no hot electrons even if $I_N$ is larger than that in Case I, which shows that the absolute growth driven by Beam-L is a must for hot electron generation in this regime. Lowering $I_L$ is proposed to be a good strategy for mitigating the hot electron fluxes in the N-L system since it does not need to alter the total input laser energy very much.

Other than the energy fluxes, the temperature ($T_{hot}$) that determines the penetration depth is another key feature of the hot electrons from the target-preheating point of view. In the simulations, $T_{hot}$ is obtained by Maxwellian fitting of the high energy tail in the electron distribution function.\cite{Yan2014} $T_{hot}$ is found to reach a quasi-steady value in the highly nonlinear stage of each case as shown in figure \ref{fig:hot}(c).  Figure \ref{fig:hot}(c) also shows that $T_{hot}$ largely depends on $I_N$ but is not that sensitive to $I_L$. In the nonlinear stage $T_{hot}$ reaches about 40 keV in Case I. Increasing $I_N$ to $5\times 10^{14} W/cm^2$ (Case II) causes $T_{hot}$ to rise to 50 keV. However, $T_{hot}$ is found very similar (about 40 keV) in Cases I and V where $I_N$ is kept the same while $I_L$ is quite different in the two cases. Staged acceleration of hot electrons in Case I is demonstrated in figure \ref{fig:hot} (d) as it shows that the maximum longitudinal momentum $p_x$ of the hot electrons increases from smaller $x$ to larger $x$ (indicated by the blue arrow). The staged acceleration process is similar to that in the single-beam normal-incidence regime reported in reference \cite{Yan2012}: The EPWs in the low-density region have low phase velocities that allow effective first-stage acceleration of thermal electrons. Then the electrons are staged accelerated by the EPWs with higher phase velocities in the higher-density region. 

The scaling laws obtained through numerical fitting of comprehensive PIC simulations \cite{Cao2022} have been reported to well predict $T_{hot}$ for the cases $
\eta \gtrsim 1$ in the single-beam normal-incidence regime. Although the two-beam N-L regime with $\eta <1$ is certainly beyond the aimed scope fo equation (2) of reference \cite{Cao2022}, this formula is still arguably adapted to predict $T_{hot}$ in an N-L regime since Beam-L can be considered as a small energy perturbation added to the single-beam normal-incidence system including Beam-N only. Equation (2) of reference \cite{Cao2022} predicts $T_{hot} = 64$ keV for Case I, larger than the simulation value $T_{hot} = 40$ keV. Better agreement is reached for larger $\eta$. Lower $T_{hot}$ poses lower preheating risk, but still, the unexpected TPD growth in the N-L regime when $\eta$ is well below unity is an uncertainty to direct-drive ICF.

\section{Summary and discussions}
To summarize, TPD driven by two laser beams in an N-L system has been investigated through PIC simulations. It is shown that in the presence of a low-intensity Beam-L that exceeds its large-angle-incidence TPD threshold ($\xi>1$), significant TPD growth is able to be driven in a regime where $\eta$ is well below unity. Both beams contribute to the growth of TPD via the ``seed-amplification" process where the absolute instability driven by Beam-L provides the seeds that get convectively amplified by Beam-N. The levels of the seeds are mostly determined by Beam-L, while the amplification factors are determined by Beam-N. The two beams contribute to the TPD growth in different ways, both making TPD energetically important and causing significant pump depletion as well as hot electron generation in the N-L system. The hot electron fluxes are found depending on the intensities of both beams while the hot electron temperature is found mostly depending on the intensity of Beam-N. 

The low threshold of TPD in this regime can be harmful to direct-drive ICF and place extra confinement on target design, as TPD is able to grow significantly in previously unexpected regions. In experimental conditions, Beam-L comes from a laser's side part that glances at the edge of the corona plasma. Therefore, to lower the risk of TPD, it would be a good strategy to shrink the width of each laser spot, which has already been well known to benefit the implosion performance by mitigating CBET.\cite{Igumenshchev2012, Igumenshchev2013b, Froula2013} Reducing the spot sizes can help lower $I_L$ in the N-L system illustrated in figure \ref{fig:scheme}(a), thereby mitigate the two-beam TPD as another pro to direct-drive ICF.

\section*{Acknowledgment}
We thank the UCLA-IST OSIRIS Consortium for the use of $OSIRIS$. This research was supported by the National Natural Science Foundation of China (NSFC) under Grant Nos. 12375243 and 12388101, by the Strategic Priority Research Program of Chinese Academy of Sciences under Grant Nos. XDA25050400 and XDA25010200, by the Science Challenge Project, and by the Fundamental Research Funds for the Central Universities. The numerical calculations in this paper have been done on the supercomputing system in the Supercomputing Center of University of Science and Technology of China.
\section*{References}
\bibliographystyle{unsrt}
\bibliography{bibNuclearVer2405}

\newpage

\begin{table*}
	\caption{PIC simulation parameters. All cases have the density scale length $L_n=100 \ \mathrm{\mu m}$ and the electron temperature $T_e = 3  \ \mathrm{keV}$. $I_{N}$ ($I_{L}$) is the intensity of Beam-N (Beam-L) and $I_{all} \equiv I_N + I_L$. All the intensities are in the unit of $10^{14}\ \mathrm{W/cm}^2$.
$\eta$ and $\xi$ are the threshold parameters of Beam-N and Beam-L calculated via equation (1) and (2), respectively. $\eta_{all}$ is calculated by using $I_{all}$ in Eq.(\ref{eq:simon-eta}). $\bar{\alpha}_{R}$ is the energy flux carried by the $\geq 50 \ \mathrm{keV}$ electrons reaching the right boundary normalized to the incident laser energy flux, averaged between $t = 5 \ \mathrm{ps}$ and $6 \ \mathrm{ps}$.}
	\begin{center}
		\begin{tabular}{ c c c c c c c c c }
			\hline 
			Index & $I_N$ & $I_L $ & $I_{all}$ & $\eta$ & $\xi$ & $\eta_{all}$ & $\bar{\alpha}_{R}$ \\
			\hline
			
			\uppercase\expandafter{\romannumeral1}    & 3.0  & 0.5   & 3.5       & 0.43      & 13.0 & 0.50 & $1.8\%$\\
			\uppercase\expandafter{\romannumeral2}    & 5.0  & 0.5   & 5.5       & 0.71      & 13.0 & 0.79 & $5.9\%$\\
			\uppercase\expandafter{\romannumeral3}    & 3.5  & 0.0   & 3.5       & 0.50      & 0.0 & 0.50 & $0.0\%$\\
			\uppercase\expandafter{\romannumeral4}    & 3.0  & 0.3   & 3.3       & 0.43      & 7.8 & 0.47 & $0.2\%$\\
			\uppercase\expandafter{\romannumeral5}    & 3.0  & 1.0   & 4.0       & 0.43      & 26.0 & 0.57 & $5.0\%$\\
			\uppercase\expandafter{\romannumeral6}(Fixed Ions)    & 3.0(after 2 ps)   & 0.5(0-2 $\mathrm{ps}$)   & 3.5       & 0.43      & 13.0 & 0.50 & $0.2 \%$\\
			\uppercase\expandafter{\romannumeral7}(Fixed Ions)    & 3.0  & 0.5   & 3.5       & 0.43      & 13.0 & 0.50 & $2.0\%$\\
			
			\hline\hline
		\end{tabular}
		\label{tab:param}
	\end{center}
\end{table*}

%

\end{document}